\documentstyle[12pt]{article}
\newlength{\extraspace}
\newlength{\extraspaces}
\newcommand{\ba}{\begin{eqnarray}
\addtolength{\abovedisplayskip}{\extraspaces}
\addtolength{\belowdisplayskip}{\extraspaces}
\addtolength{\abovedisplayshortskip}{\extraspace}
\addtolength{\belowdisplayshortskip}{\extraspace}}
\newcommand{\ea}{\end{eqnarray}}
\begin{document}

\thispagestyle{empty}

\hfill \parbox{3.5cm}{hep-th/0312127 \\ SIT-LP-03/12}
\vspace*{1cm}
\vspace*{1cm}
\begin{center}
{\large \bf Nonlinear Supersymmetric General Relativity  \\
and   \\
Unity of Nature}
\footnote{Based upon the talk given at the 3rd International Conference on  Quantum Theory and Symmetry,  
September 9-14, Cincinnati University, Ohio, USA}    \\[20mm]
{Kazunari SHIMA} \\[2mm]
{\em Laboratory of Physics, Saitama Institute of Technology}
\footnote{e-mail:shima@sit.ac.jp}\\
{\em Okabe-machi, Saitama 369-0293, Japan}\\[2mm]
{December 2003}\\[15mm]


\begin{abstract}
Nonlinear supersymmetric(NLSUSY) general relativity(GR) is considered and 
a new fundamental action of the vacuum Einstein-Hilbert(EH)-type is obtained 
by the Einstein gravity analogue geomtrical arguments 
on new spacetime inspired by NLSUSY. 
The new action of NLSUSY GR is unstable and breaks spontaneously to 
EH action with matter(or New spacetime is unstable and
 induces spontaneously the phase transition to Riemann spacetime with matter), 
which may be the real shape of the big bang of the universe and 
give naturally the unified description of spacetime and matter.
Some implications for the low energy particle physics and the cosmology are discussed.  

\end{abstract}
\end{center}

\newpage
\section{Introduction}
NLSUSY GR is imagined by the two different arguments on the (local) supersymmetry(SUSY)
\cite{wz}\cite{va}\cite{gl}  
and supergravity theory\cite{fnf}\cite{dz}. 
One is group theoretical and the other is geometrical arguments.  
SUSY is recognized as the most promissing notion for the unification of spacetime and matter. 
%
%
%
%
%
%
From the viewpoints of simplicity and beauty of nature 
the unified theory should accommodate all observed  particles in a single
irreducible representation of a certain algebra(group) especially in the case of 
spacetime having a certain boundary.   
%
%
Facing so many fundamental elementary particles 
(more than 160 for SUSY grand unified theories(GUTs)) and  arbitrary coupling parameters, 
we are tempted to suppose that they may be certain composites  and/or that 
they should be attributed to the  particular geometrical structure of spacetime. 
We have found group theoretically that among the massless irreducible representations 
of all SO(N) super-Poincar\'e(SP) groups 
N=10 SP group is the only one that contains the strong-electroweak standard model(SM) 
with just three genarations of quarks and leptons, where we have decomposed 10 supercharges into  
${\underline 10 = \underline 5 + \underline 5^{*}}$ with respect to SU(5). 
Regarding 10 {\it supercharges} as the hypothetical fudamental spin 1/2 {\it particles(superons)}-quintet 
and anti-quintet, we have proposed the composite superon-graviton model(SGM) for nature\cite{ks0}, 
where all (observed) elementary  particles except graviton are assumed as the superon-composites  and are 
{\em eigenstates}, in the sence that they can be recasted algebraiclly 
into the equivalent local fields explicitly, of SO(10) SP symmetry of nature. 
This group theoretical argument indicates the {\it field-current(charge) identity}, 
i.e. N=10 NLSUSY Volkov-Akulov(VA) model~\cite{va} in curved spacetime. 
While, the geometrical arguments are on the classical solutions of spin ${3 \over 2}$ gravitino 
in flat spacetime of SUGRA. 
We have found that the nontrivial solution of gravitino exists  on flat spacetime not of the Schwarzshild 
back ground spacetime but of the Kerr back ground spacetime\cite{ks1}. 
And we have observed that the dimensional constant $a$ of the angular momentum( {\it global} spin) 
of the Kerr solution plays a crucial role for the existence of the nontrivial (localized) solution and 
gives {\it spacetime a nontrivial topological structure},  
where two Minkowski sheets are connected through a disk of radius $a$\cite{ks1}.
(The solution vanishes as ${a \rightarrow 0}$.) 
This simple  geometrical result suggests that  SUSY in curved spacetime allows 
the nontrivial topological structure of spacetime(even flat space), i.e. not only the angular momentum 
degrees of freedom(d.o.f.) but also the local spin d.o.f. with a dimensional constant may be embedded 
into {\it flat sapcetime} d.o.f..
%
%
%
%
From this viewpoint the gravitational interraction of  NL SUSY VA model~\cite{va} 
which treats the Minkowski and the Grassman coordinates on an equal footing is an interesting problem. 
Further the NLSUSY in curved spacetime is also interesting from the mass hierarchy viewpoints, 
for it contains naturally two mass scales, one is the tiny cosmological constant from NLSUSY and 
the other is the huge Planck mass provided the gravitational interaction of VA model is constructed.
From these viewpoints we study  the gravitational interaction of spin 1/2 VA model, i.e. 
NLSUSY in curved  (Riemann) spacetime and formulate the problem as  NLSUSY GR.
\section{ Fundamental action for  NLSUSY General Relativity }
A nonlinear supersymmetric general relativity theory(NLSUSY GRT) 
(or N=1 SGM action from the composite model viewpoints) is proposed.  
We extend the geometrical arguments of Einstein general relativity theory(EGRT) on Riemann spacetime 
to new  (SGM) spacetime posessing ${\it locally}$   NL SUSY d.o.f, 
i.e.   besides the ordinary SO(3,1) Minkowski coordinate $x^{a}$ 
the SL(2C) Grassman coordinates $\psi$ for the coset space ${superGL(4,R) \over GL(4,R)}$ 
turning subsequently to the NG fermion dynamical d.o.f. are attached at every curved spacetime point. 
Note that SO(3,1) and SL(2C) are locally holomorphic {\it non-compact groups for spacetime (coordinates) d.o.f.}, 
which may be analogous to SO(3) and SU(2) {\it compact groups for gauge (fields) d.o.f.} of 
't Hooft-Polyakov monopole\cite{th}\cite{p}.  
We have obtained  the following NLSUSY GRT(N=1 SGM) action\cite{ks2} of the vacuum EH-type.
\begin{equation}
L(w)=-{c^{4} \over 16{\pi}G}\vert w \vert(\Omega + \Lambda ),
\label{SGM}
\end{equation}
\begin{equation}
\vert w \vert=det{w^{a}}_{\mu}=det({e^{a}}_{\mu}+ {t^{a}}_{\mu}(\psi)),  \quad
{t^{a}}_{\mu}(\psi)={\kappa^{4}  \over 2i}(\bar{\psi}\gamma^{a}
\partial_{\mu}{\psi}
- \partial_{\mu}{\bar{\psi}}\gamma^{a}{\psi}),
\label{w}
\end{equation} 
where $w^{a}{_\mu}(x)$ is the unified vierbein of SGM spacetime, 
G is the gravitational constant, ${\kappa^{4} = ({c^{4}\Lambda \over 16{\pi}G}})^{-1}$ 
is a fundamental volume of four dimensional spacetime of VA model~\cite{va},  
and $\Lambda$ is a  ${small}$ cosmological constant related to the strength of 
the superon-vacuum coupling constant. 
Therefore SGM contains two mass scales,  ${1 \over {\sqrt G}}$(Planck scale) in the first term describing 
the curvature energy  and $\kappa \sim {\Lambda \over G}(O(1))$ in the second term describing the 
vacuum energy of SGM, which are responsible for the masss hierarchy.
$e^{a}{_\mu}$ is the ordinary vierbein of EGRT  describing the local SO(3,1) d.o.f 
and  ${t^{a}}_{\mu}(\psi)$ itself is not the vierbein but the mimic vierbein analogue composed of 
the stress-energy-momentum tensor of superons describing the local SL(2C) d.o.f..
$\Omega$ is a new scalar curvature analogous to the Ricci scalar curvature $R$ of EGRT, 
whose explicit expression is obtained  by just replacing ${e^{a}}_{\mu}(x)$  
by ${w^{a}}_{\mu}(x)$ in Ricci scalar $R$~\cite{st1}.    \\
These results can be understood intuitively by observing that 
${w^{a}}_{\mu}(x) ={e^{a}}_{\mu}+ {t^{a}}_{\mu}(\psi)$  inspired  by 
$\omega^{a}=dx^{a} + {\kappa^{4}  \over 2i}(\bar{\psi}\gamma^{a}
d{\psi}
- d{\bar{\psi}}\gamma^{a}{\psi})
\sim {w^{a}}_{\mu}dx^{\mu}$, where $\omega^{a}$ is the NLSUSY invariant differential forms of 
VA\cite{va}, is invertible, i.e.,
\begin{equation}
w^{\mu}{_a}= e^{\mu}{_a}- t{^{\mu}}_a + {t^{\mu}}_{\rho}{t^{\rho}}_a 
- t{^{\mu}}_{\sigma} t{^{\sigma}}_{\rho}   t{^{\rho}}_a 
+t{^{\mu}}_{\kappa} t{^{\kappa}}_{\sigma}t{^{\sigma}}_{\rho}t{^{\rho}}_a + \cdots, 
\label{w-inverse}
\end{equation} 
which terminates with $(t)^{4}$ and $s_{\mu\nu} \equiv w^{a}{_\mu}\eta_{ab}w^{b}{_\nu}$ and 
$s^{\mu \nu}(x) \equiv w^{\mu}{_{a}}(x) w^{{\nu}{a}}(x)$ 
are a unified vierbein and a unified metric tensor of NLSUSY GRT in SGM spacetime\cite{{ks2},{st1}}. 
It is straightforward to show 
${w_{a}}^{\mu} w_{{\mu}{b}} = \eta_{ab}$,  $s_{\mu \nu}{w_{a}}^{\mu} {w_{b}}^{\mu}= \eta_{ab}$, ..etc. 
As read out in (\ref{w}), in contrast with EGRT we must be careful with 
{\it the order of the indices of the tensor in NLSUSY GRT}, 
i.e. the first and the second index of $w$ (and $t$)  represent those of the $\gamma$-matrix 
and the derivative on $\psi$, respectively.  
It seems natural that the ordinary vierbein $e^{a}{_\mu}$ and the mimic vierbein $t^{a}{_\mu}$ of 
the stress-enery-momentum tensor of superon are alined and contribute equally 
to the unified vierbein  $w^{a}{_\mu}$, 
i.e. to the curvature (total energy) of unified SGM spacetime, 
where the  fundamental  action of NLSUSY GRT is the  vacuum (empty space)  action of EH-type.        \\
The NLSUSY GR action  (\ref{SGM}) is invariant at least under the following transformations\cite{st2};
the following new NLSUSY transformation 
\begin{equation}
\delta^{NL} \psi ={1 \over \kappa^{2}} \zeta + 
i \kappa^{2} (\bar{\zeta}{\gamma}^{\rho}\psi) \partial_{\rho}\psi,
\quad
\delta^{NL} {e^{a}}_{\mu} = i \kappa^{2} (\bar{\zeta}{\gamma}^{\rho}\psi)\partial_{[\rho} {e^{a}}_{\mu]},
\label{newsusy}
\end{equation} 
where $\zeta$ is a constant spinor and  $\partial_{[\rho} {e^{a}}_{\mu]} = 
\partial_{\rho}{e^{a}}_{\mu}-\partial_{\mu}{e^{a}}_{\rho}$, \\
the following GL(4R) transformations due to (\ref{newsusy})  
\begin{equation}
\delta_{\zeta} {w^{a}}_{\mu} = \xi^{\nu} \partial_{\nu}{w^{a}}_{\mu} + \partial_{\mu} \xi^{\nu} {w^{a}}_{\nu}, 
\quad
\delta_{\zeta} s_{\mu\nu} = \xi^{\kappa} \partial_{\kappa}s_{\mu\nu} +  
\partial_{\mu} \xi^{\kappa} s_{\kappa\nu} 
+ \partial_{\nu} \xi^{\kappa} s_{\mu\kappa}, 
\label{newgl4r}
\end{equation} 
where  $\xi^{\rho}=i \kappa^{2} (\bar{\zeta}{\gamma}^{\rho}\psi)$, 
and the following local Lorentz transformation on $w{^a}_{\mu}$ 
\begin{equation}
\delta_L w{^a}_{\mu}
= \epsilon{^a}_b w{^b}_{\mu}
\label{Lrw}
\end{equation}
with the local  parameter
$\epsilon_{ab} = (1/2) \epsilon_{[ab]}(x)$    
or equivalently on  $\psi$ and $e{^a}_{\mu}$
\begin{equation}
\delta_L \psi = - {i \over 2} \epsilon_{ab}
      \sigma^{ab} \psi,     \quad
\delta_L {e^{a}}_{\mu} = \epsilon{^a}_b e{^b}_{\mu}
      + {\kappa^{4} \over 4} \varepsilon^{abcd}
      \bar{\psi}\gamma_5 \gamma_d \psi
      (\partial_{\mu} \epsilon_{bc}).
\label{newlorentz}
\end{equation}
The local Lorentz transformation forms a closed algebra, for example, on $e{^a}_{\mu}$ 
\begin{equation}
[\delta_{L_{1}}, \delta_{L_{2}}] e{^a}_{\mu}
= \beta{^a}_b e{^b}_{\mu}
+ {\kappa^{4} \over 4} \varepsilon^{abcd} \bar{\psi}
\gamma_5 \gamma_d \psi
(\partial_{\mu} \beta_{bc}),
\label{comLr1/2}
\end{equation}
where $\beta_{ab}=-\beta_{ba}$ is defined by
$\beta_{ab} = \epsilon_{2ac}\epsilon{_1}{^c}_{b} -  \epsilon_{2bc}\epsilon{_1}{^c}_{a}$.
The commutators of two new NLSUSY transformations (\ref{newsusy})  on $\psi$ and  ${e^{a}}_{\mu}$ 
are GL(4R), i.e. new NLSUSY (\ref{newsusy}) is the square-root of GL(4R); 
\begin{equation}
[\delta_{\zeta_1}, \delta_{\zeta_2}] \psi
= \Xi^{\mu} \partial_{\mu} \psi,
\quad
[\delta_{\zeta_1}, \delta_{\zeta_2}] e{^a}_{\mu}
= \Xi^{\rho} \partial_{\rho} e{^a}_{\mu}
+ e{^a}_{\rho} \partial_{\mu} \Xi^{\rho},
\label{com1/2-e}
\end{equation}
where 
$\Xi^{\mu} = 2i\kappa (\bar{\zeta}_2 \gamma^{\mu} \zeta_1)
      - \xi_1^{\rho} \xi_2^{\sigma} e{_a}^{\mu}
      (\partial_{[\rho} e{^a}_{\sigma]})$.
They show the closure of the algebra. 
The ordinary local GL(4R) invariance is trivial by the construction.   
Besides these familiar and intended symmetries, the unified vierbein $w^{a}{_\mu}$, therefore SGM action, 
is invariant under the following local spinor translation(ST) with the local spinor paremeter $\theta(x)$; 
${\delta \psi=\theta}$,    
$\delta e^{a}{_\mu}=
-i \kappa^{2}( \bar\theta\gamma^{a} \partial_{\mu}\psi+\bar\psi\gamma^{a} \partial_{\mu}\theta )$. 
The commutators vanish identically.
Note that the NG fermion d.o.f. $\psi$ can be transformed(redefined) away 
neither by this local ST, 
in fact, ${w(e + \delta e, t(\psi + \delta \psi))=w(e + t(\psi), 0)=w(e,t(\psi))}$ under ${\theta(x)=-\psi(x)}$ 
as indicated ${\delta w^{a}{_\mu}(x)=0}$ nor by the ordinary general coordinate transformation 
$\delta_{GL(4R)}e^{a}{_\mu}$ as far as  NLSUSY spacetime is preserved.  
Taking $\psi=0$ by any means makes SGM another theory(EH theory) based upon another flat space(Minkowski space). 
This local spinor coordinate translation invariance is somewhat puzzling(immature) so far 
but is the origin of (or recasted as) the local spinor {\it gauge} symmetry 
of the  {\it linear SUSY  gauge field} theory (SUGRA-analogue to be obtained by the linearization) 
which is equivalent to SGM  and the mass generation through the spontaneous symmetry breakdown\cite{dz2}.
%

%
%
%
%
%
%
%
Now NLSUSY GRT action (\ref{SGM}) is invariant at least under the following spacetime symmetries\cite{st2}
\begin{equation}
[{\rm new \ NLSUSY}] \otimes [{\rm local\ GL(4,R)}] 
\otimes [{\rm local\ Lorentz }] \otimes [{\rm local \ ST}],  
\label{sgmspsymm}
\end{equation}
which is isomorphic to N=1 SP group of SUGRA. 
The extension to N=10, i.e. SO(10) SP is straightforward by taking ${\psi^{j}, (j=1,2,..10)}$\cite{ks2}.   \\
As for the internal summetry we just mention that  ${w^{a}{_\mu}}$ is invariant 
under the local U(1) transformation $\psi{_j} \rightarrow e^{i \lambda_{j}(x)}\psi_{j}$ 
due to the Grassman(Majorana spinor) nature $(\psi_{j})^{2}=0$, i.e. 
(\ref{SGM}) with N-extension is invariant at least under 
\begin{equation}
[{global \ SO(N)}] \otimes [local \ U(1)]^{N}.  
\label{sgmisymm}
\end{equation}
Therefore the action (\ref{SGM}) describes the vacuum energy(everything) of the ultimate spacetime and 
is NLSUSY GRT, a nontrivial generalization of the EH action.   
It should be noticed that SGM action  (\ref{SGM}) posesses two types of flat space which are 
not equivalent, i.e. SGM-flat($w{^a}_{\mu}(x) \rightarrow {\delta}{^a}_{\mu}$)  and 
Riemann-flat($e{^a}_{\mu}(x) \rightarrow {\delta}{^a}_{\mu}$). 
As discussed later this structure plays impotant roles in the spontaneous breakdown of spacetime and 
in the cosmology of SGM (\ref{SGM}).   
%
%
%
\section{Discussions}
The linearization of NLSUSY GRT(N=1 SGM) action (\ref{SGM}), i.e. the construction (identification) of 
the renormalizable and local LSUSY {\it gauge} field theory which is equivalent to (\ref{SGM}), 
is inevitable to derive the SM as the low energy effective theory. 
Particularly N=10 must be linearized to test the composite SGM scenario, 
though some characterictic and accessible predictions, e.g. 
nutrino- and  quark-mxings, proton stability, CP violation, the generation structure.. etc. 
are obtained qualitatively by the group theoretical arguments\cite{ks0}\cite{ks2}. 
The linearizations of  N=1 and N=2  NLSUSY VA action in flat spacetime  have been carried out explicitly 
by the systematic arguments and show  that they are equivalent to the LSUSY actions 
with Fayet-Iliopoulos terms for the scalar (or axial vector) supermultiplet\cite{ik}\cite{r}
and the vector supermultiplet\cite{stt1}, respectively. 
These exact results obtained systematically as the representations of the symmetries  
by {\em the algebraic arguments} are favourable and encoraging towards 
the linearization of SGM and the (composite) SGM scenario. 
We anticipate that the local spacetime  symmetries  of SGM mentioned above  plays  crucial roles 
in the linearization, especially in constructing the SUSY invariant relations\cite{{ik},{r}}.     \\
We regard that the ultimate real shape of nature is  high symmetric new(SGM) spacetime inspired by NLSUSY, 
where the coset space coordinates $\psi$ of ${superGL(4,R) \over GL(4,R)}$ 
turning to the NG fermion d.o.f. in addition to the ordinaly Minkowski coordinate $x^{a}$, 
i.e.  $ local \ SL(2C) \times local \ SO(3,1)$ d.o.f., are attached at every  spacetime point.  
The geometry of new spacetime is described by SGM action (\ref{SGM}) of {\it vacuum \ EH-type} 
and gives the unified description of nature. 
As proved for EH action of GR\cite{wttn}, the energy of NLSUSY GR action of 
EH-type is anticipated to be positive ( $\Lambda>0$).  
NLSUSY GR action (\ref{SGM}), $L(w) \sim w\Omega + w\Lambda$, on SGM spacetime is unstable  and 
induces {\it the spontaneous (symmetry) breakdown} into EH action 
with NG fermion (massless superon-quintet) matter, 
${L(e,\psi) \sim eR + e\Lambda + (\cdots \kappa,\psi \cdots)}$\cite{st1}, on ordinary Riemann spacetime, 
for the curvature-energy potential of SGM  spacetime is released into the potential of Riemann spacetime 
and the energy-momentum of superon(matter), i.e. ${w\Omega > eR}$. 
As mentioned before SGM action poseses  two different flat spaces. 
One is SGM-flat ($w{^a}_{\mu}(x) \rightarrow {\delta}{^a}_{\mu}$)  space of NLSUSY GR action $L(w)$. 
And the other is Riemann-flat ($e{^a}_{\mu}(x) \rightarrow {\delta}{^a}_{\mu}$) space of SGM action ${L(e,\psi)}$ 
which allows (generalized) NLSUSY VA action. 
This can be regarded as the phase transition of spacetime from SGM to Riemann (with NG fermion matter).
Also this may be the birth of the present expanding universe, i.e. the big bang and 
the rapid expansion (inflation) of spacetime and matter.  
%
%
%
%
%
%
And we think that the birth of the present universe by the {\it spontaneous \ breakdown} 
of SGM spacetime  described  by {\it vacuum} action of EH-type (\ref{SGM}) 
may explain qualitatively the observed critical value$(\sim 1)$  of the energy density 
of the universe.  
%
%
%
%
%
%
%
%
%
%
It is interesting if SGM could give new insights into the unsolved problems 
of the cosmology, e.g. the origins(real shapes) of the  big bang, inflation, dark energy,
matter-antimatter asymmetry, $\cdots$, etc.   \par
In this study we have attempted a {\it geometrical} unification of spacetime and matter.
New (SGM) spacetime is the ultimate physical entity and specified by NLSUSY GRT (SGM action) (\ref{SGM})  
of vacuum EH-type. 
The study of the vacuum structure of SGM action in the broken phase
(i.e. NLSUSY GRT action in Riemann spacetime with matter) is important 
for linearizing SGM and to obtain the equivalent local LSUSY gauge field theory.            \par
SGM  with the extra dimensions, which  can be constructed straightforwardly and 
gives another unification framework by regarding the observed particles as elementary, is open.  
In this case there are two mechanisms for the conversion of the spacetime d.o.f. 
into the dynamical d.o.f., i.e. by the compactification of Kaluza-Klein type and 
by the new mechanism presented in SGM.  \par
By extending the arguments to spin 3/2 NG fermion~\cite{b}  we have obtained SGM with spin 3/2 fermion~\cite{st3}. 
%
%
%
%
%
\newpage

%
\newcommand{\NP}[1]{{\it Nucl.\ Phys.\ }{\bf #1}}
\newcommand{\PL}[1]{{\it Phys.\ Lett.\ }{\bf #1}}
\newcommand{\CMP}[1]{{\it Commun.\ Math.\ Phys.\ }{\bf #1}}
\newcommand{\MPL}[1]{{\it Mod.\ Phys.\ Lett.\ }{\bf #1}}
\newcommand{\IJMP}[1]{{\it Int.\ J. Mod.\ Phys.\ }{\bf #1}}
\newcommand{\PR}[1]{{\it Phys.\ Rev.\ }{\bf #1}}
\newcommand{\PRL}[1]{{\it Phys.\ Rev.\ Lett.\ }{\bf #1}}
\newcommand{\PTP}[1]{{\it Prog.\ Theor.\ Phys.\ }{\bf #1}}
\newcommand{\PTPS}[1]{{\it Prog.\ Theor.\ Phys.\ Suppl.\ }{\bf #1}}
\newcommand{\AP}[1]{{\it Ann.\ Phys.\ }{\bf #1}}


\begin{thebibliography}{100}
%
\bibitem{wz} J. Wess and B. Zumino, {\it Phys. Lett.} {\bf B49}, 52 (1974).  
%
\bibitem{va}  D.V. Volkov and V.P. Akulov, {\it Phys. Lett.} {\bf B46}, 109(1973). 
%
\bibitem{gl}  Y.A. Golfand and E.S. Likhtman, {\it JET. Lett.} {\bf 13}, 323 (1971).
%
\bibitem{fnf}  D.Z. Freedman, P. van Nieuwenhuizen and S. Ferrara, {\it Phys. Rev.} {\bf D13}, 3214(1976). 
%
\bibitem{dz}  S. Deser and B. Zumino, {\it Phys. Lett.}{\bf B62} (1976) 335.
%
\bibitem{ks0}  K. Shima, {\it European. Phys. J.} {\bf C7}, 341(1999).
%
\bibitem{ks1} K. Shima and M. Kasuya, {\it Phys. Rev.} {\bf D22}, 290(1980). 
%
\bibitem{th}  G. 't Hooft,  {\it Nucl. Phys.} {\bf B79}, 276(1974).
%
\bibitem{p}  A. M. Polyakov,  {\it JETP. Lett.} {\bf 20}, 194(1974).
%
\bibitem{ks2}  K. Shima,  {\it Phys. Lett.} {\bf B501}, 237(2001).
%
\bibitem{st1}  K. Shima and M. Tsuda, {\it Class. and Quantum Grav.} {\bf 19}, 1 (2002).
%
\bibitem{st2}  K. Shima and M. Tsuda, {\it Phys. Lett.} {\bf B507}, 260(2001).
%
\bibitem{dz2}  S. Deser and B. Zumino,  {\it Phys. Rev. Lett.} {\bf 38} (1977) 1433.
%
\bibitem{ik}   E. A. Ivanov and A.A. Kapustnikov, {\it J. Phys.} {\bf A11}, 2375(1978).
%
\bibitem{r}    M. Ro\v{c}ek, {\it Phys. Rev. Lett.} {\bf 41}, 451(1978).
%
\bibitem{stt1}   K. Shima, Y. Tanii and M. Tsuda, {\it Phys. Lett.} {\bf B546}, 162 (2002).
%
\bibitem{wttn} E. Witten,  {\it Commun. Math. Phys.}, {\bf80}, 381(1981).
%
\bibitem{b}  N.S. Baaklini, {\it Phys. Lett.} {\bf 67B}, 335(1976).
%
\bibitem{st3} K. Shima and M. Tsuda, {\it Phys. Lett.} {\bf B521}, 67(2001).
%
\end{thebibliography}
\end{document}